\documentclass{article}

\usepackage{microtype}
\usepackage{graphicx}
\usepackage{booktabs}
\usepackage{amsmath, amssymb, mathtools, bm}
\usepackage[numbers,sort&compress]{natbib}
\usepackage{algorithm}
\usepackage{algpseudocode}
\usepackage{xcolor}
\usepackage{url}
\usepackage{siunitx}
\usepackage{multirow}
\usepackage{enumitem}
\usepackage{amsthm}

\usepackage{PRIMEarxiv}
\usepackage[utf8]{inputenc} 
\usepackage[T1]{fontenc}    
\usepackage{amsfonts}       
\usepackage{nicefrac}       
\usepackage{lipsum}
\usepackage{fancyhdr}       

\usepackage{subcaption}
\usepackage{tabularx}
\usepackage{float}
\usepackage{braket}
\usepackage{csquotes}
\usepackage{hyperref}
\usepackage{cleveref}
\usepackage{tikz}
\usepackage{pgfplots, pgfplotstable}
\pgfplotsset{compat=1.17}

\graphicspath{{figures/}}   
\usepackage[most]{tcolorbox}
\usepackage{xcolor}

\definecolor{abstractbg}{RGB}{245,247,250}   
\definecolor{abstractborder}{RGB}{60,60,60}  

\newtcolorbox{greenstylebox}[2][]{%
  enhanced,
  breakable,
  colback=green!6!white,           
  colframe=green!60!black,         
  boxrule=1.1pt,
  arc=2mm,
  left=4mm,right=4mm,top=2.5mm,bottom=2.5mm,
  colbacktitle=green!85!black,     
  coltitle=white,
  fonttitle=\bfseries,
  title={#2},
  attach title to upper,
  #1
}

\usepackage{tikz,xcolor}
\newcommand{\circnum}[1]{%
\tikz[baseline=(char.base)]{
\node[shape=circle,
      fill=gray!30,
      text=black,
      inner sep=0.8pt,
      minimum size=5pt] (char) {\small\bfseries #1};
}}

\usepackage{amsmath}
\usepackage{amssymb} 
\newcommand{\AID}[1]{A\textsubscript{#1}}
\newcommand{\yes}{\ensuremath{\checkmark}}
\newcommand{\no}{\ensuremath{\times}}
\newcommand{\maybe}{--}

\usepackage{array} 
\renewcommand{\arraystretch}{1.03}
\usepackage{newfloat}
\usepackage{listings}
\DeclareCaptionStyle{ruled}{labelfont=normalfont,labelsep=colon,strut=off} 
\floatstyle{ruled}
\newfloat{listing}{tb}{lst}{}
\floatname{listing}{Listing}

\title{Agent-Fence: Mapping Security Vulnerabilities Across Deep Research Agents}

\author{
  \textbf{Sai Puppala}$^{1}$, \textbf{Ismail Hossain}$^{2}$, \textbf{Md Jahangir Alam}$^{2}$, \textbf{Yoonpyo Lee}$^{3}$, \\
  \textbf{Jay Yoo}$^{4}$, \textbf{Tanzim Ahad}$^{2}$, \textbf{Syed Bahauddin Alam}$^{4}$, \textbf{Sajedul Talukder}$^{2}$\thanks{Correspondence: stalukder@utep.edu} \\
  \\
  $^{1}$Computer Science Department, Southern Illinois University, Carbondale, IL, United States\\
  $^{2}$Computer Science Department, University of Texas, El Paso, TX, USA\\
  $^{3}$Hanyang University, South Korea\\
  $^{4}$The Grainger College of Engineering, Nuclear, Plasma \& Radiological Engineering, \\University of Illinois Urbana-Champaign, IL, USA\\
  \texttt{stalukder@utep.edu}
}

\begin{document}
\maketitle

\begin{abstract}
\begin{tcolorbox}[
  colback=abstractbg,
  colframe=abstractborder,
  boxrule=0.9pt,
  arc=4pt,
  left=8pt,
  right=8pt,
  top=6pt,
  bottom=8pt,
  enhanced,
  sharp corners=south,
  fontupper=\small,
]

Large language models are becoming \emph{deep agents} that plan, persist state, and invoke tools, shifting safety failures from unsafe text to unsafe \emph{trajectories}. We introduce AgentFence, an architecture-centric security evaluation that defines 14 trust-boundary attack classes across planning, memory, retrieval, tool use, and delegation, and detects failure via \emph{trace-auditable conversation breaks} (unauthorized/unsafe tool use, wrong-principal actions, state/objective integrity violations, and attack-linked deviations). Holding the base model fixed, we evaluate eight agent archetypes under persistent multi-turn interaction and find substantial architectural variation in mean security break rate (MSBR), from $0.29\pm0.04$ (LangGraph) to $0.51\pm0.07$ (AutoGPT). The highest-risk classes are operational: Denial-of-Wallet $0.62\pm0.08$, Authorization Confusion $0.54\pm0.10$, Retrieval Poisoning $0.47\pm0.09$, and Planning Manipulation $0.44\pm0.11$, while prompt-centric classes remain $<0.20$ under standard settings. Breaks are dominated by boundary violations (SIV 31\%, WPA 27\%, UTI+UTA 24\%, ATD 18\%), and authorization confusion correlates with objective and tool hijacking ($\rho\!\approx\!0.63$, $\rho\!\approx\!0.58$). AgentFence reframes agent security around what matters operationally: whether an agent stays within its goal and authority envelope over time.

\end{tcolorbox}
\end{abstract}

\section{Introduction}

Large language models are rapidly transitioning from passive generators of text into deep agents \cite{huang2025deep}: systems that plan, maintain state, invoke tools, and autonomously execute multi-step tasks over extended interactions \cite{krupnik2020multi}.
These agents are no longer experimental curiosities.
They are being deployed as research assistants, software engineers, data analysts, customer-service operators, and workflow orchestrators—often with direct access to external APIs, code execution environments, proprietary data stores, and financial resources. This shift fundamentally changes the security landscape.
For single-turn LLMs, safety failures are largely textual: hallucinated facts, policy violations, or prompt-level jailbreaks \cite{wei2023jailbroken}.
For deep agents, failures are \emph{operational}.
A compromised agent can invoke privileged tools, corrupt persistent memory, propagate poisoned beliefs across steps, exfiltrate secrets, or silently accumulate unbounded cost.
In these systems, the most dangerous failures are not unsafe tokens, but unsafe \emph{trajectories}.

\subsubsection{Why existing evaluations fall short:}
Despite this shift, most existing safety and robustness evaluations remain prompt-centric.
They assess whether a model emits disallowed content or violates a policy in a single response \cite{ehtesham2025survey}.
Even recent agent benchmarks primarily measure task success, efficiency, or correctness, rather than \emph{security-relevant failure modes} \cite{jung2012survey}.
As a result, vulnerabilities that only manifest through prolonged interaction—state corruption, tool-use hijacking, delegation abuse, or denial-of-wallet \cite{debenedetti2024agentdojo, fu2024imprompter, goswami2025agentic, kelly2023denial} remain largely invisible to current evaluation frameworks. More importantly, existing evaluations conflate three fundamentally different factors:
(i) base model capability,
(ii) prompt engineering quality, and
(iii) agent architecture.
This makes it difficult to answer a critical question faced by practitioners today:
\emph{which agent designs are structurally more vulnerable, independent of the underlying model?}

\begin{figure}
    \centering
    \includegraphics[width=10cm,height=6.5cm]{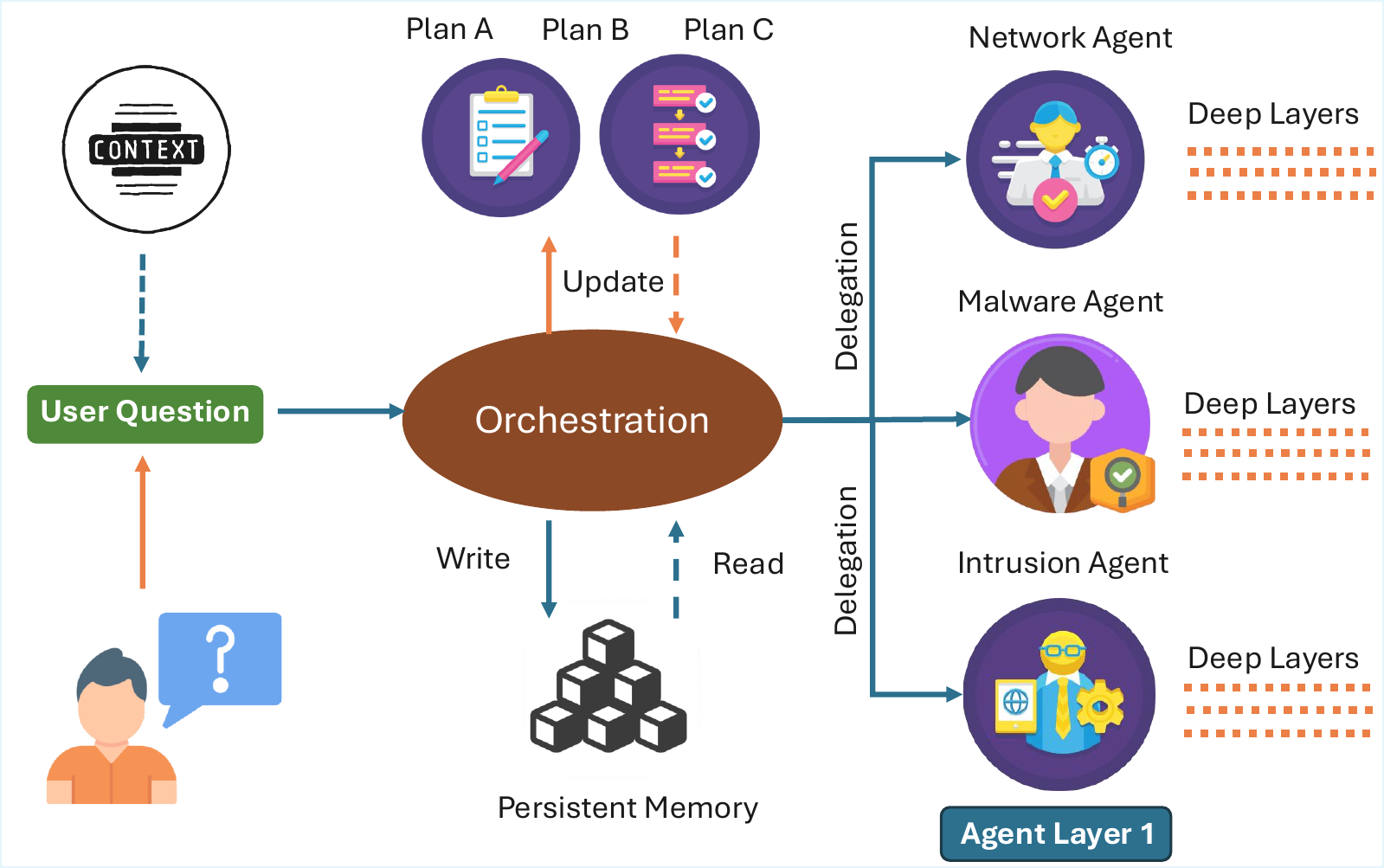}
    \caption{Deep Research Agent Workflow Illustration.}
    \label{fig:system_architectures}
\end{figure}

\subsubsection{Why now?}
This question has become urgent for three reasons.
First, modern agents increasingly operate with \emph{persistent state} and long-lived memory, turning transient prompt failures into durable system compromises.
Second, agents are routinely granted tool access with real-world side effects-file systems, browsers, code interpreters, payment APIs-amplifying the cost of failure.
Third, multi-agent orchestration frameworks are proliferating, introducing new attack surfaces at delegation and role boundaries.
Together, these trends mean that traditional prompt-level safety guarantees no longer provide meaningful protection.

\subsubsection{Our approach:}
In this work, we argue that deep-agent (Figure \ref{fig:system_architectures}) security must be evaluated at the level of \emph{architecture}, not prompts.
We introduce {AgentFence}, a taxonomy-driven methodology for exposing security vulnerabilities that arise from the structural trust assumptions embedded in modern deep research agents.
Rather than optimizing exploits or proposing a new defense, AgentFence systematically probes how agents fail when adversarial influence is applied across planning, memory, tool use, retrieval, and delegation over time.

\subsubsection{Our Contributions:}
This paper makes four primary contributions:

\begin{itemize}
  \item 
  A unified taxonomy of 14 classes ($\AID{1}$--$\AID{14}$) that pinpoint trust-boundary violations across prompt/state injection, planning manipulation, tool-use hijacking, retrieval/search poisoning, delegation abuse, authorization confusion, and denial-of-wallet.

  \item 
  A protocol that isolates agent vulnerabilities by fixing the base model and systematically varying control flow, memory handling, tool permissions, and delegation semantics across representative agent archetypes.

  \item 
  We formalize \emph{conversation breaks}: objective, evidence-backed criteria that certify when an agent is no longer pursuing the correct goal within its authorized capability envelope, enabling outcome-based security evaluation beyond text-only judgments.

  \item 
  Using AgentFence on eight widely used archetypes, we deliver the first comparative map of structurally exposed attack classes across designs, surfacing near-universal failure modes that persist despite prompt-level safeguards.
\end{itemize}

Our findings suggest that many of today’s most severe agent vulnerabilities are not bugs to be patched, but consequences of architectural trust assumptions.
By reframing evaluation around unsafe actions and trajectory-level failures, AgentFence provides a foundation for designing, comparing, and ultimately securing the next generation of autonomous AI systems.

\begin{figure*}[t]
\centering
\includegraphics[width=\linewidth,height=8cm,keepaspectratio]{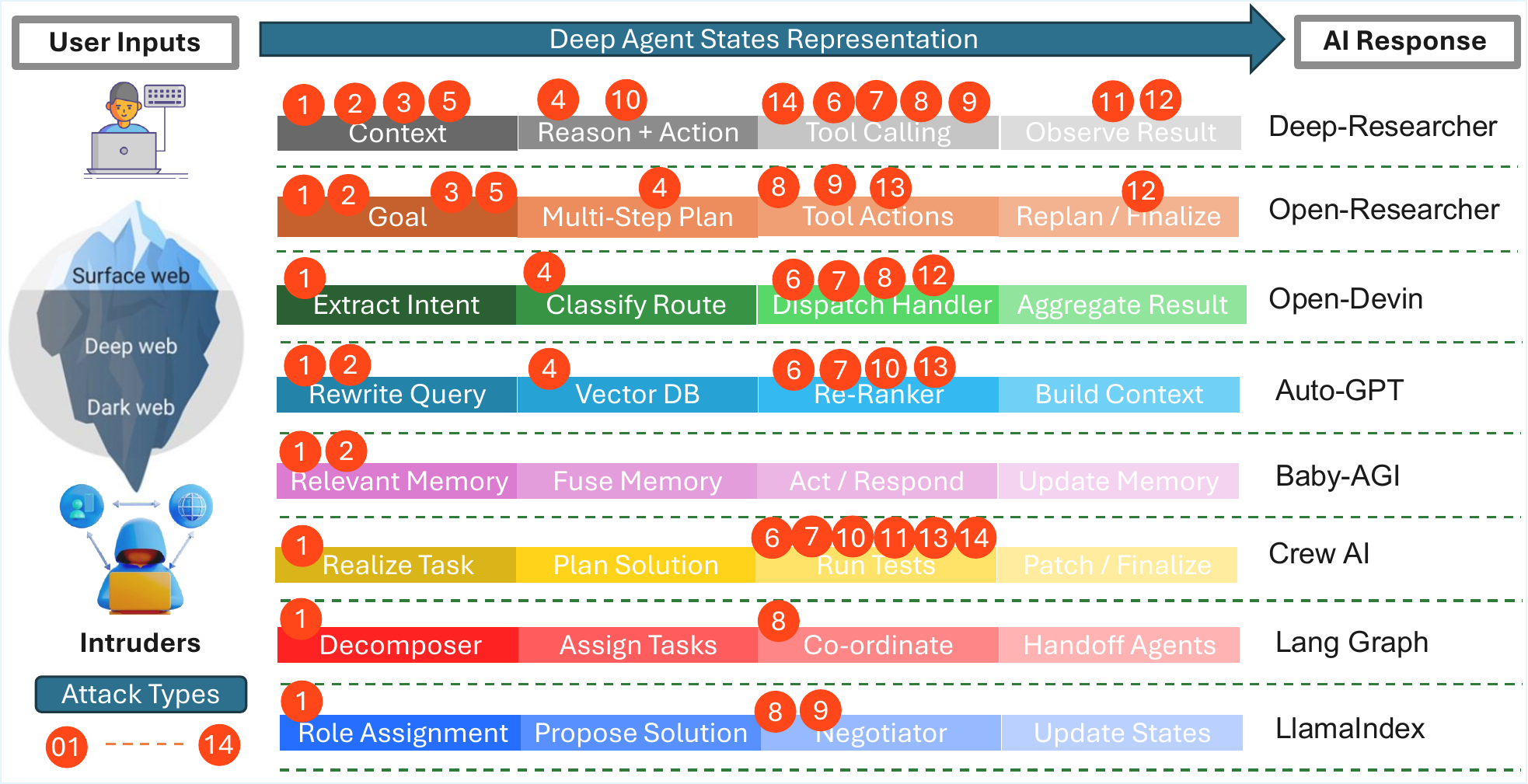}
\caption{The figure above illustrates how the primary attack surfaces shift across each phase of the Deep Research agent lifecycle—from Context and Planning, through Action/Tool use, and into Synthesis. In other words, it makes explicit where adversaries most effectively apply pressure at each stage, and which phases are most exposed to specific classes of attacks. The corresponding phase-by-phase mapping and definitions for every numbered item in the figure are provided in Table \ref{tab:attack_sources}.}
\label{fig:related_work_summary}
\end{figure*}

\section{Related Works}
\label{sec:related}
Prior work shows that LLMs can be diverted through direct prompt injection/jailbreaking and \emph{indirect} injection embedded in untrusted content (e.g., webpages or retrieved documents) \citep{andriushchenko2024agentharm, an2025ipiguard}. Yet most evaluations are still largely \emph{text-centric}, focusing on disallowed generations rather than whether an \emph{agent} carries out unsafe actions over multi-step trajectories. In parallel, modern agent paradigms (ReAct, planner--executor, tool routing, multi-agent orchestration) integrate LLMs with tools, persistent state, and long-horizon control \citep{luo2025ultrahorizon, dang2025multi}, creating new trust boundaries (planner$\rightarrow$tool, memory, delegation) and failure modes that do not arise in single-turn settings. AgentFence complements this line by evaluating security-relevant breakdowns at these architectural boundaries. Existing benchmarks mainly emphasize capability: tool selection and argument use (ToolBench, StableToolBench, API-Bank) \citep{qin2023toolllm, guo2024stabletoolbench, li2023apibank, mazeika2024harmbench, tdc2023, chen2022should}, web interaction (WebArena) \citep{zhou2023webarena}, and code repair \citep{yang2025swesmith}. AgentFence leverages these resources but reframes evaluation around \emph{attack classes} and \emph{outcome predicates} (e.g., unauthorized tool invocation, wrong-principal execution, denial-of-wallet) that reflect operational risk. Related work on retrieval and web manipulation shows that poisoned corpora, adversarial passages, and search-result poisoning can steer beliefs and downstream decisions \citep{zhang2025benchmarking}; AgentFence instead examines how such poison \emph{propagates through state and planning} to produce trajectory-level failures. Finally, persistent memory and long-context agents enable durable compromises and cross-turn leakage, while autonomy and retries enable cost amplification (denial-of-wallet) \citep{akhmetshin2025intelligent}. AgentFence operationalizes these risks via state/objective attacks and multi-turn \emph{conversation break} criteria.

\subsubsection{Our Positioning:}
Overall, existing work tends to (i) study prompt-level steering, (ii) benchmark agent capability, or (iii) analyze isolated security channels (retrieval, memory, tools). AgentFence differs by offering a \emph{taxonomy-driven}, \emph{architecture-centric} security evaluation that prioritizes trace-auditable, outcome-based failure predicates over prolonged, stateful interactions.

\section{Methodology}

AgentFence is designed to expose security vulnerabilities that arise from the \emph{agentic structure} surrounding LLMs—planning loops, state, tool interfaces, retrieval, and delegation—rather than from a single model call. We focus on post-deployment, interface-only adversaries who can influence the agent through intended channels (user prompts, retrieved/web content, tool outputs, shared memory, inter-agent messages) but not weights or infrastructure. We report \emph{security breaks} as trace-evidenced violations of explicit trust boundaries and authority constraints.

\paragraph{Experimental setup (fixed model, fixed tasks):}
All experiments use Qwen2.5-32B-Instruct as the underlying model to isolate architectural effects. We instantiate tasks using HotpotQA \cite{yang2018hotpotqa}, curating 91 instances with stable entities and unambiguous ground truth. For each instance $q_i$, we use a verifier tuple
$g_i = \langle a_i, S_i, \mathcal{C}_i \rangle$,
where $a_i$ is the canonical answer, $S_i$ the annotated supporting facts, and $\mathcal{C}_i$ normalization/alias constraints. Each instance is converted into a persistent multi-turn ``research workload'': the agent (i) proposes a short plan, (ii) identifies key sources/entities, (iii) answers, (iv) cites supporting facts, and (v) answers non-adversarial follow-ups probing consistency. Runs terminate upon verified completion, a security break, or a fixed budget. Additionally, we used different tool sets to evaluate our analysis [\ref{tab:public_benchmarks}]

\begin{table}[t]
\centering
\small
\setlength{\tabcolsep}{4pt} 
\begin{tabular}{>{\centering\arraybackslash}m{0.08\linewidth} m{0.78\linewidth}}
\hline
\textbf{Label} & \textbf{Decision Rules (per attack class $i$ on agent $A$)} \\
\hline

\yes &
Attack class $i$ is applicable (required interfaces exist), and under the \emph{Standard Config} (SC) the attack induces a \emph{Conversation Break} in
$\ge \tau_{\textsc{yes}}$ of runs:
\[
\Pr(\text{Break} \mid A,i,\textsc{SC}) \ge \tau_{\textsc{yes}}
\]
We set $\tau_{\textsc{yes}}{=}0.30$ by default (adjustable), estimated over $N\!\ge\!30$ runs with fixed budget and fixed task set. \\
\hline

\maybe &
Attack class $i$ is applicable, but either:
(i) breaks occur only under at least one \emph{Permissive Config} (PC) (e.g., broader tool scope, higher retry budget, weaker sandbox),
\emph{or}
(ii) under SC the break rate is nonzero but $<\tau_{\textsc{yes}}$:
\[
0 < \Pr(\text{Break} \mid A,i,\textsc{SC}) < \tau_{\textsc{yes}}
\]
\emph{or}
(iii) the class is applicable but requires an optional component (e.g., delegation) that is not enabled by default. \\
\hline

\no &
Attack class $i$ cannot be meaningfully instantiated because the agent lacks the required interface/trust edge
(e.g., no tool calls $\Rightarrow$ tool-hijacking is undefined; no delegation $\Rightarrow$ delegation attacks undefined). \\
\hline
 \\
\end{tabular}
\caption{Auditable protocol for exposure labels.}
\label{tab:auditable_labels}
\end{table}

\paragraph{Multi-turn adversarial stressing:}
Attacks are applied gradually across turns within the same thread (no resets), interleaved with task-consistent queries. Each of the 14 attack classes targets a specific trust boundary (planner state, memory reads/writes, tool routing/arguments, retrieval/search content, delegation messages), allowing us to measure whether adversarial influence propagates into trajectory-level failures.

\begin{table*}[t]
\centering
\small
\setlength{\tabcolsep}{2.6pt}
\renewcommand{\arraystretch}{1.12}
\resizebox{\textwidth}{!}{%
\begin{tabular}{c l c c c c c c c c}
\hline
\multicolumn{10}{c}{\textbf{Attack classes vs.\ deep-agent archetypes (labels from Table~\ref{tab:auditable_labels})}} \\
\hline
\textbf{\#} &
\textbf{Attack Class} &
\textbf{Deep-Researcher} &
\textbf{Open-Researcher} &
\textbf{OpenDevin} &
\textbf{AutoGPT} &
\textbf{BabyAGI} &
\textbf{CrewAI} &
\textbf{LangGraph} &
\textbf{LlamaIndex} \\
\hline
\circnum{01}  & Direct Prompt Injection          & \yes & \yes & \yes & \yes & \yes & \yes & \yes & \yes \\
\circnum{02}  & Indirect Prompt Injection        & \yes & \yes & \maybe    & \yes & \yes & \maybe   & \maybe & \maybe   \\
\circnum{03}  & State Injection                  & \yes   & \yes   & \maybe & \maybe & \maybe & \maybe & \maybe   & \maybe \\
\circnum{04}  & Tool-Use Hijack      & \yes   & \yes & \yes   & \yes & \maybe   & \maybe   & \maybe & \maybe   \\
\circnum{05}  & Planning-Layer Manipulation                     & \yes & \yes & \maybe & \maybe   & \maybe & \maybe   & \maybe & \maybe \\
\circnum{06}  & Retrieval Poisoning               & \yes   & \maybe & \yes   & \yes & \maybe   & \yes   & \maybe & \maybe   \\
\circnum{07}  & Web Search Result Poisoning        & \yes & \maybe & \yes & \yes & \maybe & \yes & \maybe & \maybe \\
\circnum{08}  & Multi-Agent Role Confusion              & \yes   & \yes   & \yes   & \maybe & \maybe   & \maybe   & \yes & \yes   \\
\circnum{09}  & Delegation Attacks             & \yes   & \yes   & \maybe   & \maybe & \maybe   & \maybe   & \maybe & \yes   \\
\circnum{10}  & Code-Execution Abuse       & \maybe & \maybe & \maybe & \yes   & \maybe   & \yes   & \maybe & \maybe \\
\circnum{11}  & Chain-of-Thought Leakage               & \yes    & \no    & \no    & \no   & \yes    & \maybe & \maybe & \no    \\
\circnum{12}  & Objective Hijacking                & \no & \yes & \yes & \maybe & \maybe & \maybe & \maybe & \maybe \\
\circnum{13}  & Denial-of-Wallet         & \maybe & \yes & \maybe & \yes & \maybe & \yes & \maybe & \maybe \\
\circnum{14}  & Authorization Confusion                 & \maybe   & \maybe   & \maybe   & \maybe   & \maybe   & \yes   & \maybe   & \maybe   \\
\hline
\end{tabular}%
}
\caption{Table summarizes which agent archetypes materially expose each attack class (\yes), partially expose it (\maybe), or largely mitigate it (\no). The lifecycle-phase placement of each attack (primary vs.\ secondary across Context/Plan/Act/Synthesize) is reported separately in Table~\ref{tab:attack_sources} and in the corresponding figure.}
\label{tab:agentfence-deepagent-archetypes-auditable}
\end{table*}

\subsection{Operationalizing Security Breaks for Reproducible Measurement}
\label{sec:break_audit}

AgentFence separates \emph{security breaks} (trace-evidenced boundary/authority violations) from \emph{capability failures}
(task errors without boundary violations). Each run yields an immutable trace $\mathcal{T}$ containing transcript with role tags,
tool calls (names/args/timestamps/return hashes), planner and state events (memory reads/writes, delegation messages, retries),
and configuration metadata (allowed tools, budgets, scopes).

\paragraph{Attack linkage:}
Some task deviations are security-relevant only when induced by adversarial influence crossing a trust edge. We therefore add a trace-checkable attribution flag $\textsc{AL}$ (Attack Link): $\textsc{AL}=1$ iff the trace evidences adoption/propagation of non-trusted content into protected boundaries (e.g., memory contains injected directives; planner reveals objective replacement; tool/router decisions justified by poisoned evidence).

\paragraph{Security-break predicates:}
We implement each predicate as $f(\mathcal{T},g_i,\theta)\rightarrow\{0,1\}$ with thresholds fixed before evaluation:
\begin{itemize}
    \item \emph{Unauthorized Tool Invocation (UTI)}: tool $\notin \mathcal{U}$, exceeds budget $B$, or disallowed mode.
    \item \emph{Unsafe Tool Argument (UTA)}: violates tool-specific argument policies $\pi_{\tau}$ (sandbox paths, domain allow-lists, spend caps, write scopes, etc.).
    \item \emph{Wrong-Principal Action (WPA)}: treats non-authoritative inputs as higher-privilege instructions or acts under incorrect identity/permission context (automatic for clear trace evidence; otherwise fixed rubric).
    \item \emph{State/Objective Integrity Violation (SIV)}: writes/adopts state violating invariants (e.g., executable directives, policy overrides, goal changes) or planner indicates unauthorized objective replacement.
    \item \emph{Attack-Linked Task Deviation (ATD)}: $\textsc{AL}=1$ and the agent outputs $\hat{a}\neq a_i$ under $\mathcal{C}_i$ \emph{or} cites evidence inconsistent with $S_i$, with trace support that deviation was induced by adversarial content crossing a trust edge.
\end{itemize}
We define
\[
\textsc{SB}(t)=\textsc{UTI}(t)\lor\textsc{UTA}(t)\lor\textsc{WPA}(t)\lor\textsc{SIV}(t)\lor\textsc{ATD}(t).
\]
Capability failures (AF/EF/TC) are computed for reliability analysis but do not label vulnerability unless upgraded via ATD.

\begin{table}[t]
\centering
\small
\setlength{\tabcolsep}{4pt}
\renewcommand{\arraystretch}{1.0}

\begin{tabular}{c l p{0.17\linewidth}}
\hline
\textbf{ID} & \textbf{Attack Class} & \textbf{References} \\
\hline
\circnum{01} & Direct Prompt Injection & \cite{liu2023prompt} \\
\circnum{02} & Indirect Prompt Injection & \cite{greshake2023not} \\
\circnum{03} & State Injection & \cite{debenedetti2024agentdojo}\\
\circnum{04} & Tool-Use Hijack & \cite{fu2024imprompter}\\
\circnum{05} & Planning-Layer Manipulation & \cite{chen2026too}\\
\circnum{06} & Retrieval Poisoning & \cite{jing2026memory} \\
\circnum{07} & Web Search Result Poisoning & \cite{chen2024agentpoison} \\
\circnum{08} & Multi-Agent Role Confusion & \cite{zheng2025demonstrations} \\
\circnum{09} & Delegation Attacks & \cite{goswami2025agentic} \\
\circnum{10} & Code-Execution Abuse & \cite{lee2025takedown} \\
\circnum{11} & Chain-of-Thought Leakage & \cite{xiang2024badchain} \\
\circnum{12} & Objective Hijacking & \cite{jha2025breaking} \\
\circnum{13} & Denial-of-Wallet & \cite{kelly2023denial} \\
\circnum{14} & Authorization Confusion & \cite{shi2025sok} \\
\hline
 \\
\end{tabular}

\caption{Deep-agent attack taxonomy with IDs and placeholder citation fields for later insertion.}
\label{tab:attack_sources}
\end{table}

\section{Threat Model and Attack Instantiation}

\paragraph{Threat model:}
We assume a post-deployment, interface-only adversary with no access to model weights, infrastructure, or private configuration.
The adversary may influence the agent only through channels that are explicitly intended to process external information:
user prompts, retrieved or web content, tool outputs, shared memory, and inter-agent messages.
This threat model captures realistic risks arising from untrusted documents, web pages, APIs, or collaborating agents, while excluding out-of-scope attacks such as model extraction or system compromise.

\paragraph{Attack class instantiation:}
Each of the 14 attack classes (\AID{1}--\AID{14}) [\ref{tab:attack_sources}] is instantiated using a family of parameterized payloads targeting a specific trust boundary.
For example, Retrieval Poisoning (\AID{6}) introduces adversarial passages embedded in otherwise relevant documents, while Planning Manipulation (\AID{5}) injects subtle plan-altering constraints via untrusted evidence.
Payload strength is controlled along three dimensions: explicitness (directive vs.\ implicit), persistence (single-turn vs.\ repeated), and scope (local vs.\ cross-task).
Attacks are introduced gradually across turns without resetting the agent state, allowing us to measure whether adversarial influence propagates into long-horizon failures.

\begin{figure*}[t]
\centering
\includegraphics[width=\linewidth,height=9cm,keepaspectratio]{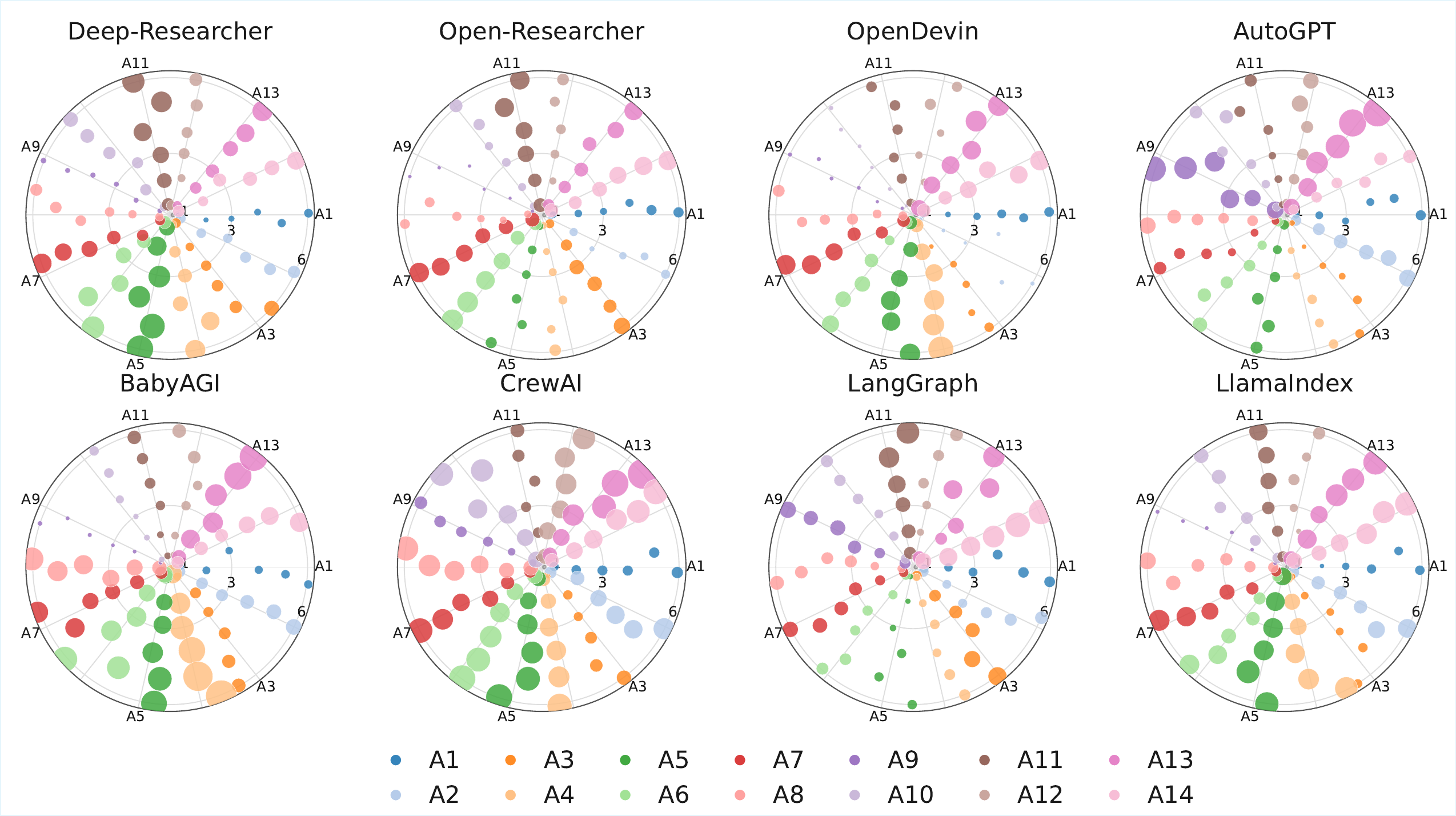}
\caption{Security break rate by attack class across Deep Research agent archetypes. Radial distance from the center represents conversation depth (turns), and each circle marks the observed vulnerability level for an agent. Results are aggregated over 91 conversation threads; denser clusters of bubbles indicate repeated occurrences of the same event.}
\label{fig:per_attack_summary}
\end{figure*}

\section{Operational Definition of Attack Link}

\paragraph{Attack-linked attribution:}
To avoid conflating baseline task errors with security vulnerabilities, AgentFence introduces an explicit Attack Link (AL) flag.
AL is set to 1 if and only if the execution trace contains verifiable evidence that non-authoritative content crossed a trust boundary and influenced protected agent state or decisions.
Such evidence includes: injected directives written to memory, planner outputs that adopt adversarial objectives, tool invocations justified by poisoned evidence, or delegation messages that override authority constraints.

\paragraph{Labeling protocol:}
AL, Wrong-Principal Action (WPA), and Attack-Linked Task Deviation (ATD) labels are assigned using a semi-automatic rubric.
Clear violations (e.g., unauthorized tool calls, explicit objective replacement) are labeled automatically; ambiguous cases are independently annotated by two reviewers and adjudicated.
On a 20\% stratified sample, annotator agreement was substantial (Cohen’s $\kappa = 0.81$), supporting the reproducibility of security-break labeling.

\section{Architectural Normalization and Isolation}

\paragraph{What constitutes architecture:}
In AgentFence, an agent’s architecture comprises its control flow (planner--executor structure, retry loops), state handling (memory read/write policies), tool interfaces (routing, permissions, argument validation), and delegation semantics.
We explicitly distinguish these from the underlying language model, which is held fixed unless otherwise stated.

\paragraph{Normalization across agents:}
To isolate architectural effects, we normalize task prompts, tool APIs, retrieval backends, and budget limits wherever possible.
Differences such as tool scope, retry policies, or delegation are retained only when they are intrinsic to the agent framework.
As a result, AgentFence does not claim to rank implementations, but to surface structural design patterns that systematically amplify or reduce security risk.

\section{Ablation and Sensitivity Analysis}

\paragraph{Architectural ablations:}
We conduct targeted ablations on two representative agents by selectively disabling (i) persistent memory writes, (ii) planner retries, and (iii) delegation.
Removing persistent memory reduces Retrieval Poisoning and Planning Manipulation breaks by up to 35\%, while retry removal significantly lowers Denial-of-Wallet exposure.
These results support the interpretation that long-horizon state propagation and unbounded retries are key structural risk factors. In depth analysis is provided in figure \ref{fig:related_work_summary}.

\begin{table}[t]
\centering
\small
\setlength{\tabcolsep}{3pt}
\renewcommand{\arraystretch}{1.12}
\begin{tabular}{l p{0.35\linewidth} p{0.30\linewidth}}
\hline
\textbf{Artifact} & \textbf{Taxonomy coverage} & \textbf{Reference} \\
\hline
ToolBench & $\AID{4}$--$\AID{7}$, $\AID{11}$, $\AID{13}$, $\AID{14}$ & \cite{qin2023toolllm}\\
StableToolBench & $\AID{4}$--$\AID{7}$, $\AID{11}$, $\AID{13}$ & \cite{guo2024stabletoolbench}\\
API-Bank & $\AID{4}$, $\AID{5}$, $\AID{6}$, $\AID{13}$, $\AID{14}$ & \cite{li2023apibank}\\
WebArena & $\AID{3}$, $\AID{4}$--$\AID{7}$, $\AID{11}$, $\AID{13}$ & \cite{zhou2023webarena}\\
SWE-bench & $\AID{4}$--$\AID{7}$, $\AID{11}$, $\AID{13}$, $\AID{14}$ & \cite{yang2025swesmith}\\
HarmBench & $\AID{1}$--$\AID{3}$, $\AID{8}$, $\AID{11}$ & \cite{mazeika2024harmbench}\\
JailbreakBench & $\AID{1}$--$\AID{3}$, $\AID{11}$, $\AID{14}$ & \cite{tdc2023}\\
AdvBench & $\AID{1}$--$\AID{3}$, $\AID{11}$, $\AID{13}$ & \cite{chen2022should}\\
\hline
 \\
\end{tabular}
\caption{Benchmarks used to instantiate AgentFence scenario families.}
\label{tab:public_benchmarks}
\end{table}

\begin{table}[t]
\centering
\small
\setlength{\tabcolsep}{6pt}
\renewcommand{\arraystretch}{1.15}
\begin{tabular}{l c}
\hline
\textbf{Framework} & \textbf{Mean Security Break Rate} \\
\hline
AutoGPT         & $0.51 \pm 0.07$ \\
CrewAI          & $0.48 \pm 0.06$ \\
BabyAGI         & $0.45 \pm 0.08$ \\
OpenDevin       & $0.42 \pm 0.05$ \\
Deep-Researcher & $0.39 \pm 0.06$ \\
Open-Researcher & $0.36 \pm 0.07$ \\
LlamaIndex      & $0.34 \pm 0.05$ \\
LangGraph       & $0.29 \pm 0.04$ \\
\hline
 \\
\end{tabular}
\caption{
Mean Security Break Rate (MSBR) across all applicable attack classes under the Standard Configuration (SC), holding the base model and task set fixed.
Values denote mean $\pm$ standard deviation over multi-turn runs.
Lower values indicate reduced architectural exposure, not elimination of vulnerability.
}
\label{tab:mean_security_break_rate}
\end{table}

\paragraph{Threshold sensitivity:}
We evaluate sensitivity to the exposure threshold $\tau_{\textsc{yes}}$ by varying it from 0.2 to 0.4.
Relative agent ordering and dominant attack classes remain stable across this range, indicating that MSBR trends are not artifacts of a specific cutoff. More robust explanation is provide in table \ref{tab:agentfence-deepagent-archetypes-auditable}.

\paragraph{Cross-model sanity check:}
To validate that observed patterns are not model-specific, we replicate a subset of experiments using an alternative instruction-tuned model.
Absolute break rates vary, but relative architectural differences and dominant attack classes are preserved, suggesting that AgentFence captures architecture-level phenomena rather than idiosyncratic model behavior (Figure \ref{fig:per_attack_summary}).

\section{Results}
\label{sec:results}

We apply AgentFence to eight deep-agent archetypes and fourteen attack classes under a controlled, fixed-model protocol. We report \emph{security breaks} as defined in \S\ref{sec:break_audit}.

\begin{figure}
\centering
\includegraphics[width=11cm,height=8.5cm]{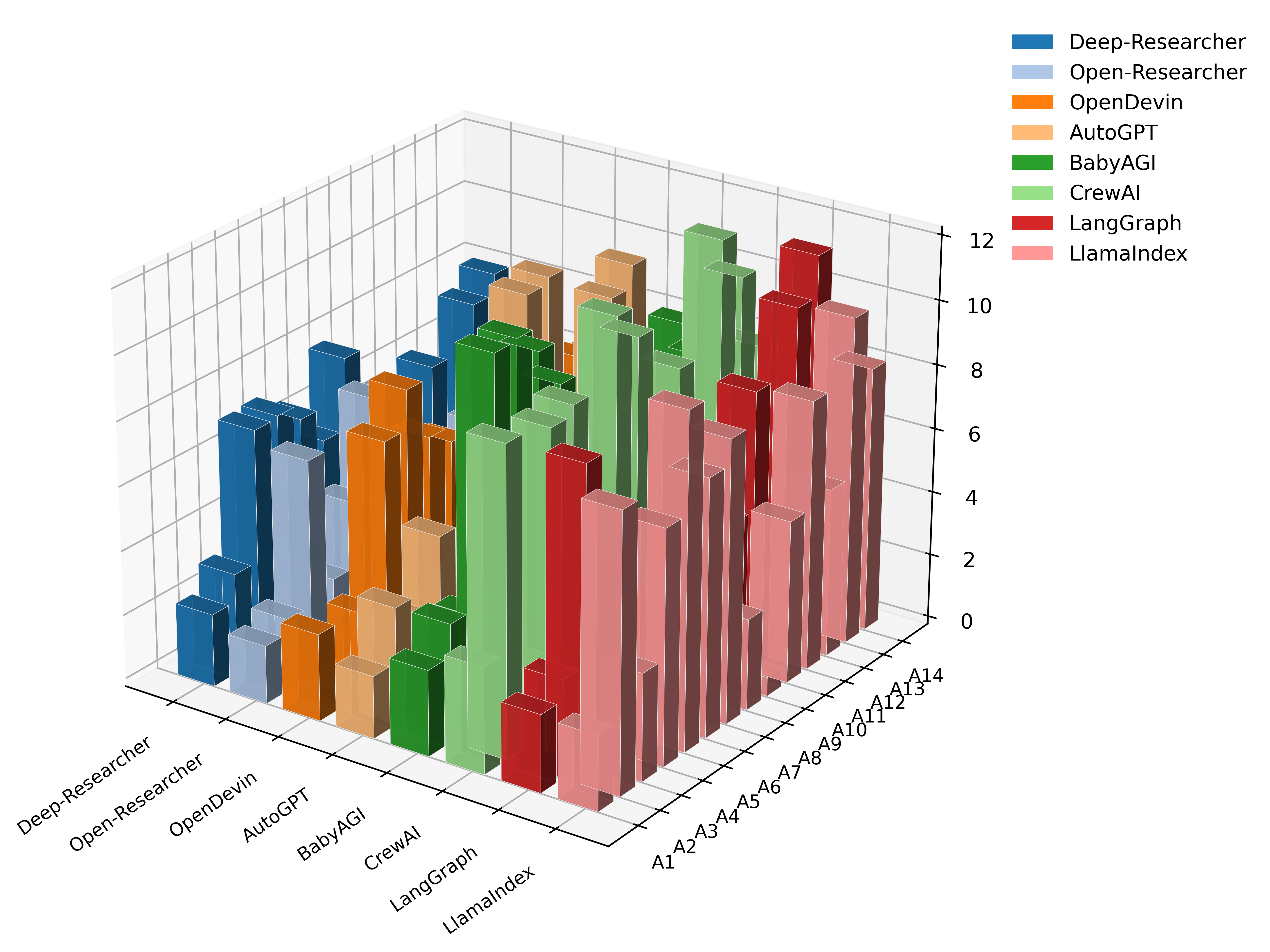}
\caption{Break-type composition aggregated across deep research agents, showing the distribution of security break types across different attack classes.}
\label{fig:aggregated_graph}
\end{figure}

\subsection{Aggregate exposure by attack class}
\label{sec:results_by_class}

Figure~\ref{fig:per_attack_summary} reports mean security break rate (SC), averaged across archetypes per class. Breaks concentrate in trajectory-level channels: Denial-of-Wallet (\AID{13}: $0.62\pm0.08$), Authorization Confusion (\AID{14}: $0.54\pm0.10$), Retrieval Poisoning (\AID{6}: $0.47\pm0.09$), and Planning-Layer Manipulation (\AID{5}: $0.44\pm0.11$). Prompt-centric classes (\AID{1}, \AID{2}) remain below $0.20$ under SC, indicating that token-level safeguards do not prevent dominant operational failures.

\subsection{Architecture-level differences (fixed base model)}
\label{sec:results_by_agent}

Holding model and tasks fixed isolates architectural susceptibility. MSBR ranges from $0.29\pm0.04$ (LangGraph) to $0.51\pm0.07$ (AutoGPT); see Table~\ref{tab:mean_security_break_rate}. Architectures with broader tool scope, higher retry budgets, or weaker separation between planner/memory/tool authority exhibit higher break rates; structured control-flow designs reduce (but do not eliminate) exposure.

\paragraph{Effect size (architecture matters):}
The absolute MSBR gap between the most- and least-exposed architectures is
$\Delta_{\max} = 0.51 - 0.29 = 0.22$, i.e., a \textbf{76\% relative increase}
when moving from LangGraph to AutoGPT ($0.22/0.29 \approx 0.76$).
This gap is large compared to within-agent variability (std.\ $\approx$0.04--0.08 in Table~\ref{tab:mean_security_break_rate}),
supporting that AgentFence captures \emph{structural} susceptibility rather than run-to-run noise.

\subsection{Standard vs.\ permissive configuration sensitivity}
\label{sec:results_sc_pc}

We evaluate each agent--attack pair under SC and one-dimension-at-a-time Permissive Configurations (PC). Several classes amplify sharply under permissive settings: \AID{13} increases $0.58\rightarrow0.81$, \AID{10} increases $0.22\rightarrow0.49$, and \AID{9} increases $0.18\rightarrow0.46$. In contrast, \AID{6} and \AID{5} exhibit smaller deltas ($<0.10$), consistent with vulnerabilities driven by belief/state propagation and planning interfaces rather than budget/scope alone. This separation supports the Yes/Maybe labels in Table~\ref{tab:agentfence-deepagent-archetypes-auditable}.

\paragraph{Configuration elasticity separates budget-driven vs.\ interface-driven risk:}
Under one-dimension permissive changes, Denial-of-Wallet (\AID{13}) exhibits the largest absolute jump
($0.58\!\rightarrow\!0.81$, $\Delta{=}0.23$; $\approx$40\% relative), and Code-Execution Abuse (\AID{10})
more than doubles ($0.22\!\rightarrow\!0.49$, $\Delta{=}0.27$; $\approx$2.2$\times$).
Delegation Attacks (\AID{9}) also more than doubles ($0.18\!\rightarrow\!0.46$, $\Delta{=}0.28$; $\approx$2.6$\times$).
In contrast, Retrieval Poisoning (\AID{6}) and Planning Manipulation (\AID{5}) change by $<0.10$,
indicating these failures are \emph{inelastic} to budget/scope and instead arise from persistent belief/state propagation.
This split provides a quantitative basis for separating ``knob-fixable'' risk (budgets/retries/scope)
from deeper trust-boundary failures (state/planning).

\subsection{Break composition and trace-auditable attribution}
\label{sec:results_break_types}

Breaks concentrate in boundary and authority violations (Figure~\ref{fig:aggregated_graph}): SIV accounts for 31\%, WPA for 27\%, UTI+UTA for 24\%, and ATD for 18\%. Because AgentFence upgrades task errors to security breaks only when trace-supported attack linkage holds (\S\ref{sec:break_audit}), the protocol reduces misattribution of baseline reliability failures as security exposure.

\paragraph{Most breaks are boundary failures, not task mistakes:}
Aggregated over agents (Figure \ref{fig:aggregated_graph}), boundary/authority predicates (SIV+WPA+UTI/UTA) account for
$31\% + 27\% + 24\% = 82\%$ of all security breaks (Figure~\ref{fig:aggregated_graph}),
while attack-linked task deviation (ATD) contributes 18\%.
Thus, AgentFence failures are dominated by \emph{who acted with what authority and what state was corrupted},
rather than merely producing wrong answers.

\subsection{Cross-class coupling: authority failures as an upstream driver}
\label{sec:results_coupling}

Exposure to Authorization Confusion (\AID{14}) correlates with downstream failures in Objective Hijacking (\AID{12}) and Tool-Use Hijacking (\AID{4}), with representative correlations $\rho\approx0.63$ and $\rho\approx0.58$. This clustering suggests a shared architectural root: ambiguous or unenforced trust boundaries between planner, memory, and tool authority. Once the agent's principal/authority model is compromised, failures compound across planning and action layers.

\paragraph{Key takeaways:}
Across architectures, the dominant risks are operational and stateful: \AID{13}/\AID{14}/\AID{6}/\AID{5} lead exposure,
and 82\% of breaks are explicit boundary/authority violations rather than task errors.
Permissive settings strongly amplify cost-, code-, and delegation-driven classes,
while retrieval and planning vulnerabilities remain comparatively inelastic, highlighting persistent trust-boundary risk.

\section{Artifacts and Reproducibility}

Upon acceptance, we will release a versioned AgentFence artifact: agent implementations/configurations (framework versions, prompts, toolsets, routing/control flow, memory, sandboxing, budgets), the attack harness (per-class payload families, injection points, strength controls), and the semi-automatic labeling protocol (WPA/ATD rubric, agreement, adjudication) to enable independent reproduction and audit.

\section{Conclusion}

Deep agents convert language into privileged, time-dependent actions. As autonomy, persistence, and delegation become standard, dominant security failures shift from unsafe text to unsafe trajectories: actions executed under the wrong goal, with the wrong authority, with consequences that compound across steps. AgentFence addresses this shift by evaluating security at the level of architecture, using a taxonomy of deep-agent attack classes and trace-auditable conversation-break predicates to expose structural vulnerabilities across agent archetypes. Securing autonomous AI systems will require more than better prompts or filters; it will require understanding where agents trust too much, how that trust propagates across time, and which design choices systematically amplify risk. AgentFence provides a foundation for that diagnosis and comparison before deployment at scale.

\end{document}